\documentclass[12pt,english]{article}
\usepackage[T1]{fontenc}
\usepackage[latin9]{inputenc}
\usepackage{geometry}
\usepackage{verbatim}
\usepackage{units}
\usepackage{amsmath}
\usepackage{amssymb}
\usepackage{cancel}
\usepackage{graphicx}
\usepackage{wasysym}
\usepackage{hyperref}
\makeatletter

\providecommand{\tabularnewline}{\\}

\newenvironment{lyxcode}
	{\par\begin{list}{}{
		\setlength{\rightmargin}{\leftmargin}
		\setlength{\listparindent}{0pt}
		\raggedright
		\setlength{\itemsep}{0pt}
		\setlength{\parsep}{0pt}
		\normalfont\ttfamily}%
	 \item[]}
	{\end{list}}

\usepackage{soul}

\makeatother

\usepackage{babel}
\begin{document}

\title{A universality class for RNA-like polymers and double polymers}
\author{R. Dengler\thanks{ORCID: 0000-0001-6706-8550}}
\maketitle

\begin{abstract}
We examine the statistics of conformations of a linear polymer in
a solvent. The polymer is allowed to form double polymers. We closely
follow a classical technique to derive a field theory for the problem
from an $O\left(n\right)$ symmetric spin model. The field theory
is a model for RNA or DNA with constant binding energy per monomer. 

It is shown that there is a stable renormalization group fixed point,
at which the double polymer decouples from the single-strand polymer
and becomes a branched polymer of the conventional type with a three-point
interaction. To reach this fixed point, at least one parameter must
be adjusted. The critical dimension is eight. Fisher-renormalization,
equation of state and critical exponents are reproduced in this limit.
The single-strand polymer depends on the double-strand polymer and
disappears at the critical point, but has its own critical exponents.
\end{abstract}

\section*{Introduction }

The statistical physics of linear polymers, branched polymers and
RNA-like polymer networks in a solvent has a long history. Classical
results are a method to calculate the statistics of polymer configurations
with path integrals \cite{Edwards1965} and the mapping of the configurations
of a linear polymer to the spin configurations of $O\left(n\right)$-symmetric
magnetic systems in the $n\rightarrow0$ limit \cite{Gennes1972}.

According to this $O\left(n\right)$-correspondence, the radius of
gyration of a linear polymer of length $s$ grows like $s^{\nu}$
for long polymers, where the exponent $\nu$ is universal and greater
than $\nicefrac{1}{2}$ when polymer-polymer repulsion (excluded volume
effect) is taken into account. The excluded volume effect is irrelevant
for linear polymers above four dimensions, where $\nu=\nicefrac{1}{2}$.

The limit $n\rightarrow0$ of a special $O\left(n\right)$-symmetric
spin system also was used by Lubensky and Isaacson \cite{Lubensky1978,Lubensky1979}
to examine branched polymers, that is, polymer configurations containing
two-valent and multi-valent (star like) monomers.

These authors examine connections to percolation, lattice animals,
gelation, and the Potts model. There also exists a mapping of the
critical properties of such polymer networks to the Lee-Yang edge
singularity of a spin system in $d-2$ dimensions \cite{Parisi1981}.
A related result is a mapping to a hard-core gas in $d-2$ dimensions
\cite{Brydges2003,Cardy2003}.

At a more formal level, the authors find a critical dimension $8$
when excluded volume effects are taken into account and a critical
dimension $6$ in a theta solvent (without polymer-polymer repulsion).

More complicated branched polymers, and in particular RNA-like conformations,
have been examined with many different methods \cite{Gennes1968,Mueller2007,David2009,R.Everaers2017,Rosa2019,S.Cocco2022},
only to cite a few. The irregular and not completely random base sequence
of RNA plays a significant role in most realistic cases. 

RNA molecules with a periodic base sequence like ATAT... constitute
a more tractable problem, and it suggests itself to start with this
case \cite{Gennes1968}. It turns out that the techniques used by
Lubensky and Isaacson for simple branched polymers are directly applicable
to this problem.

We follow this path and start with a rather explicit derivation of
a spin model analogous to Ref. \cite{Lubensky1979}. The examination
of the resulting field theory then in principle is standard, but as
for one-component branched polymers, the ,,theory is very rich``
\cite{Lubensky1979}.

\section*{Partition sum}

The starting point is an expression for the partition sum $Z$ of
the system, essentially the number of polymer configurations in a
lattice model. The expression makes use of a probability distribution
$P_{N}\left(\sigma_{1},\sigma_{2},...,\sigma_{N}\right)\mathrm{d}^{N}\sigma$
for $N$ spin variables $\sigma$ with the property $\left\langle \sigma_{\alpha}\sigma_{\beta}\right\rangle =\delta_{\alpha,\beta}$
and $\left\langle \sigma_{\alpha_{1}}...\sigma_{\alpha_{k}}\right\rangle =O\left(N^{k/2-1}\right)$
for $k>2$, see appendix A \cite{Gennes1979,Lubensky1979}. Expectation
values with an odd power of any $\sigma_{\alpha}$ vanish. Examples
are
\begin{align*}
\left\langle \sigma_{1}^{2}\right\rangle  & =1,\\
\left\langle \sigma_{1}^{4}\right\rangle  & =\frac{3N}{N+2},\qquad\qquad\qquad\left\langle \sigma_{1}^{2}\sigma_{2}^{2}\right\rangle =\frac{N}{N+2},\\
\left\langle \sigma_{1}^{6}\right\rangle  & =\frac{15N^{2}}{\left(N+2\right)\left(N+4\right)},\qquad\left\langle \sigma_{1}^{2}\sigma_{2}^{2}\sigma_{3}^{2}\right\rangle =\frac{N^{2}}{\left(N+2\right)\left(N+4\right)}.
\end{align*}
The essential point is that in the limit $N\rightarrow0$ there only
remains $\left\langle 1\right\rangle =1$ and $\left\langle \sigma_{\alpha}\sigma_{\beta}\right\rangle =\delta_{\alpha,\beta}.$ 

We now consider a rectangular lattice with $n$ spins $s^{\mu}$ and
$\tfrac{1}{2}n\left(n+1\right)$ spins $S^{\mu\nu}=S^{\nu\mu}$ with
$1\leq\mu,\nu\leq n$ at each lattice point. Spins $s^{\mu}$ can
be identified with the single-polymer, spins $S^{\mu\nu}$ with the
double-polymer. For each lattice point we use the probability distribution
$P_{N}\left(s^{\mu},S^{\mu\mu}/\sqrt{2},\left(S^{\mu\nu}\right)_{\mu<\nu}\right)$
for $N=n+\tfrac{1}{2}n\left(n+1\right)$ spins. This leads to
\begin{align*}
\left\langle s^{\mu}s^{\nu}\right\rangle  & =\delta_{\mu\nu},\qquad\qquad\left\langle S^{\mu\nu}S^{\rho\tau}\right\rangle =\delta_{\mu\rho}\delta_{\nu\tau}+\delta_{\mu\tau}\delta_{\nu\rho},
\end{align*}
and all other correlation functions vanish in the limit $n\rightarrow0$.
The partition sum is given by 
\[
Z=\left\langle \prod_{i,j;\mu}\left(1+\tfrac{1}{2}s_{i}^{\mu}v_{i,j}s_{j}^{\mu}\right)\prod_{i,j;\mu,\nu}\left(1+\tfrac{1}{2}S_{i}^{\mu\nu}v_{i,j}S_{j}^{\mu\nu}\right)\prod_{i}\left(1+f\left(s_{i},S_{i}\right)\right)\right\rangle ,
\]
where $i$ and $j$ denote lattice points and $v_{i,j}=1$ for next
neighbors and $v_{i,j}=0$ otherwise. The function $f$ is the interaction. 
To understand the connection between $Z$ and polymer configurations
consider next neighbor pair $i,j$ of the first product. To this pair
there corresponds the factor

\[
z_{i,j}=\prod_{\mu}\left(1+X_{ij}^{\mu}\right)=1+\sum_{\mu}X_{ij}^{\mu}+\tfrac{1}{2}\sum_{\mu,\nu}X_{ij}^{\mu}X_{ij}^{\nu}-\tfrac{1}{2}\sum_{\mu}X_{ij}^{\mu}X_{ij}^{\mu}+O\left(\left(X_{ij}\right)^{3}\right)\rightarrow\left(1+\sum_{\mu}s_{i}^{\mu}s_{j}^{\mu}\right)
\]
where $X_{ij}^{\mu}=s_{i}^{\mu}s_{j}^{\mu}+\tfrac{1}{4}s_{i}^{\mu}s_{i}^{\mu}s_{j}^{\mu}s_{j}^{\mu}$,
and in the last step all terms not contributing to the $\left\langle ...\right\rangle $
average in the $n\rightarrow0$ limit have been omitted. The remaining
,,links`` $\sum_{\mu}s_{i}^{\mu}s_{j}^{\mu}$ must be paired in
$Z$, and generate all self-avoiding paths with constant $\mu$ along
a path (a real path also requires an external source and sink).

The double polymer spin $S$ in the second product of $Z$ has two
indices, but can be decoded in the same way. It counts all self-avoiding
double polymer paths twice, with index pair $\mu\nu$ along a path
conserved or exchanged. The alternatives are caused by $S^{\mu\nu}=S^{\nu\mu}$
and can be interpreted as direct and twisted connections. In fact,
because of $\left\langle s_{i}^{2}S_{i}^{2}\right\rangle =0$ single
and double polymer paths avoid each other mutually.

The interaction term $f$ resembles the interaction in the lattice
model of one-component branched polymers \cite{Lubensky1979} and
is described in appendix B. The transition from $Z$ to a field theory
is standard and can be performed as in ref. \cite{Lubensky1979}.

\section*{Action integral}

\begin{figure}
\centering{}\includegraphics[scale=1.2]{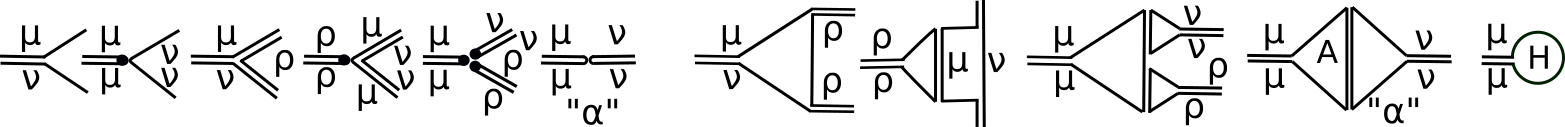}\caption{Left:\label{fig:Spin_Interactions} Fundamental and generated interactions.
The letters are field indices, the dot symbolizes $\bar{\chi}=\sum_{\mu}\chi_{\mu\mu}$.
Right: The fundamental interaction generates other interactions. The
bilinear term $A$ and the linear term $H$ play a special role.}
\end{figure}
The action integral can be written in the form

\begin{align}
S & =\int\mathrm{d}^{d}x\left\{ -\tfrac{1}{2}\sum\varphi_{\mu}\left(r_{0}-\nabla^{2}\right)\varphi_{\mu}-\tfrac{1}{2}\sum\chi_{\mu\nu}\left(\tau_{0}-\nabla^{2}\right)\chi_{\mu\nu}+g\sum_{\mu\nu}\chi_{\mu\nu}\varphi_{\mu}\varphi_{\nu}+g'\bar{\chi}\sum_{\mu}\varphi_{\mu}\varphi_{\mu}\right\} \label{eq:ActionIntegral}\\
 & +\int\mathrm{d}^{d}x\left\{ \lambda\sum_{\mu\nu\rho}\chi_{\mu\nu}\chi_{\nu\rho}\chi_{\rho\mu}-\tfrac{\alpha}{2}\bar{\chi}^{2}+\lambda'\bar{\chi}\sum_{\mu\nu}\chi_{\mu\nu}^{2}+\lambda''\bar{\chi}^{3}\right\} -H\int\mathrm{d}^{d}x\bar{\chi}\nonumber \\
 & +\int\mathrm{d}^{d}x\left\{ -u_{1}\sum_{\mu\nu}\varphi_{\mu}^{2}\varphi_{\nu}^{2}-u_{2}\sum\chi_{\mu\nu}^{2}\chi_{\rho\tau}^{2}-u_{3}\sum\varphi_{\mu}^{2}\chi_{\rho\tau}^{2}+w_{2}'\sum\chi_{\mu\nu}^{2}\chi_{\rho\tau}^{2}\chi_{\alpha\beta}^{2}+...\right\} .\nonumber
\end{align}
The field $\varphi_{\mu}$ denotes the single-strand polymer, the
field $\chi_{\mu\nu}=\chi_{\nu\mu}$ the double-strand polymer, $\bar{\chi}$
is an abbreviation for $\bar{\chi}=\sum_{\mu}\chi_{\mu\mu}.$  The
indices run from $1$ to $n$, at the end one takes the limit $n\rightarrow0$.
The limit eliminates all types of closed loops. What graphically looks
like a $\chi$-loop may still give a contribution because of $\chi$-twists,
an example is graph $F'$ in fig.(\ref{fig:SpinModel_1Loop}). 

The lattice model originally generates the interactions with coupling
constants $g$, $u_{i}$ and $w_{i}$. The $g$-interaction, however,
generates other interactions, see fig.(\ref{fig:Spin_Interactions}).
It will turn out that $g'$, $\lambda'$ and $\lambda''$ decouple
and can be ignored, but the linear term with coupling constant $H$
and the bilinear term with coupling constant $\alpha$ play a special
role. The diagrams $A$ and $H$ in fig.(\ref{fig:Spin_Interactions})
generate such contributions with originally $\alpha<0$.

The $\chi$-Propagator can be found by inverting the matrix $A_{\mu\nu;\rho\tau}$
in the $\chi$-$\chi$ two-form of $S$. For wave vectors $k$ and
$p$ and with the abbreviation $v_{k}=\tau_{0}+k^{2}$ one finds

\begin{align*}
\left\langle \chi_{k}^{\mu\nu}\chi_{p}^{\rho\tau}\right\rangle  & =\left(\frac{\delta_{\mu\rho}\delta_{\nu\tau}+\delta_{\mu\tau}\delta_{\nu\rho}}{2v_{k}}-\frac{\alpha\delta_{\mu\nu}\delta_{\rho\tau}}{v_{k}\left(v_{k}+n\alpha\right)}\right)\left(2\pi\right)^{d}\delta^{d}\left(k+p\right),\\
\left\langle \chi_{k}^{\mu\nu}\bar{\chi}_{p}\right\rangle  & =\left(\frac{1}{v_{k}}-\frac{\alpha n}{v_{k}\left(v_{k}+n\alpha\right)}\right)\delta_{\mu\nu}\left(2\pi\right)^{d}\delta^{d}\left(k+p\right),\\
\left\langle \bar{\chi}_{k}\bar{\chi}_{p}\right\rangle  & =\left(\frac{n}{v_{k}}-\frac{\alpha n^{2}}{v_{k}\left(v_{k}+n\alpha\right)}\right)\left(2\pi\right)^{d}\delta^{d}\left(k+p\right).
\end{align*}
The second and third lines are contractions of the first line over
$\rho\tau$ and $\mu\nu$.

The $\left\langle \chi\chi\right\rangle $-propagator is analogous
to the propagator of simple polymer networks \cite{Lubensky1979,Parisi1981},
and it is expedient to think of \emph{two} propagators - a less singular
normal propagator ($\sim v_{k}^{-1}$) and more singular propagator
with $\alpha$-insertions ($\sim-\alpha v_{k}^{-2}$ for $n=0$).
The $\alpha$-propagator is graphically denoted with a bar, see fig.(\ref{fig:SpinModel_1Loop}).
To get the most singular contributions to a given vertex function
one has to take the $\alpha$-propagator a maximum number of times
\cite{Lubensky1979,Parisi1981}.

\subsection*{Fisher-renormalization of critical exponents}

Hairpin diagrams like $H$ in fig.(\ref{fig:Spin_Interactions}) and
$H'$ in fig.(\ref{fig:SpinModel_1Loop}) generate a linear contribution
to $S$ like $-H\int\mathrm{d}^{d}x\bar{\chi}$ with a \emph{negative}
or positive (when $\alpha$ is positive) $H$. This term is strongly
relevant and must be included in the action integral. As usual one
shifts the field $\chi_{\mu\nu}\rightarrow\delta_{\mu\nu}Q+\chi_{\mu\nu}$,
$\bar{\chi}\rightarrow nQ+\bar{\chi}$ to eliminate the linear term\cite{Lubensky1979,Parisi1981}.
For $n\rightarrow0$ this leads to
\begin{align}
S\left(H,\tau_{0},r_{0}\right) & \rightarrow S\left(0,\tau_{0}-6\lambda Q,r_{0}-2gQ\right)\label{eq:Q_Def}\\
 & +\int\mathrm{d}^{d}x\left(\left[-H-\tau_{0}Q+3\lambda Q^{2}\right]\bar{\chi}-4u_{2}Q^{2}\bar{\chi}^{2}-4u_{2}Q\bar{\chi}\sum\chi_{\mu\nu}^{2}+...\right).\nonumber 
\end{align}
One could imagine that $H=H\left(\tau_{0}\right)$ depends on $\tau_{0}$
in such a way that $\left[...\right]\bar{\chi}$ vanishes for some
constant $Q=Q_{c}$ in all orders of the loop expansion. One then
has a simplified version of the field theory with shifted temperature
variables $\tau_{0}$ and $r_{0}$ and constant $Q.$ Linear terms
can be ignored, the contributions proportional to $u_{2}Q$ and $u_{2}Q^{2}$
do not change the structure of the field theory (except when $\alpha$
originally is negative and now becomes positive). The $\chi$-sector
of this field theory, as will be seen below, is related to the Lee-Yang
edge singularity in $d-2$ dimensions.

Actually, however, $H$ is constant, and the linear term only vanishes
if $Q$ depends on $\tau_{0}.$ A rather generic Ansatz is
\begin{equation}
Q=Q\left(\tau_{0}\right)\cong Q_{c}+a\tau-A\tau^{\beta_{H}},\label{eq:EqOfState}
\end{equation}
where $\tau=\tau_{0}-\tau_{c}$ and $\beta_{H}$ is some exponent.
For this $Q\left(\tau_{0}\right)$ the field theory has the same structure
as for $Q=\mathrm{const}$, but with a temperature variable $O\left(\tau\right)+6\lambda A\tau^{\beta_{H}}$
instead of $\tau_{0}$ (the variable $r_{0}-2gQ$ still is linear
in $r_{0}$).

If $\beta_{H}<1$, then $\tau^{\beta_{H}}$ dominates $\tau$. The
critical exponents $\nu_{H}$ and $\gamma_{H}$ of the $\chi$-field
thus are Fisher-renormalized \cite{Fisher1968} versions of the $Q=\mathrm{const}$
exponents $\nu$ and $\gamma$ ($\eta$ is not renormalized) 
\[
\nu_{H}=\beta_{H}\nu,\qquad\gamma_{H}=\beta_{H}\gamma.
\]
The correlation lenght, for instance is $\xi\propto\tau^{-\nu_{H}}$
instead of $\xi\propto\tau^{-\nu}.$ All this is analogous to simple
polymer networks \cite{Lubensky1979}. In tree-approximation the
linear term vanishes for%
{} 
\begin{align*}
Q\left(\tau_{0}\right) & =\tfrac{1}{6\lambda}\left(\tau_{0}-\sqrt{\tau_{0}^{2}-12\lambda H}\right)=\tfrac{1}{6\lambda}\left(\tau_{0}-\sqrt{\left(\tau_{0}-\tau_{c}\right)\left(\tau_{0}+\tau_{c}\right)}\right),\\
\tau' & =\tau_{0}-6\lambda Q=\sqrt{\left(\tau_{0}-\tau_{c}\right)\left(\tau_{0}+\tau_{c}\right)}.
\end{align*}
There remains the model without the linear term, but with a temperature
variable $\tau'\sim\sqrt{\left(\tau_{0}-\tau_{c}\right)}$ instead
of $\tau_{0}$. In tree approximation the critical exponents for the
$\chi$ fields thus are $\nu_{H}=\tfrac{1}{4}$, $\gamma_{H}=\tfrac{1}{4}$
and $\beta_{H}=\tfrac{1}{2}$ \cite{ZimmStock}.

\section*{Renormalization group fixed points}
\begin{figure}
\centering{}\includegraphics[scale=0.25]{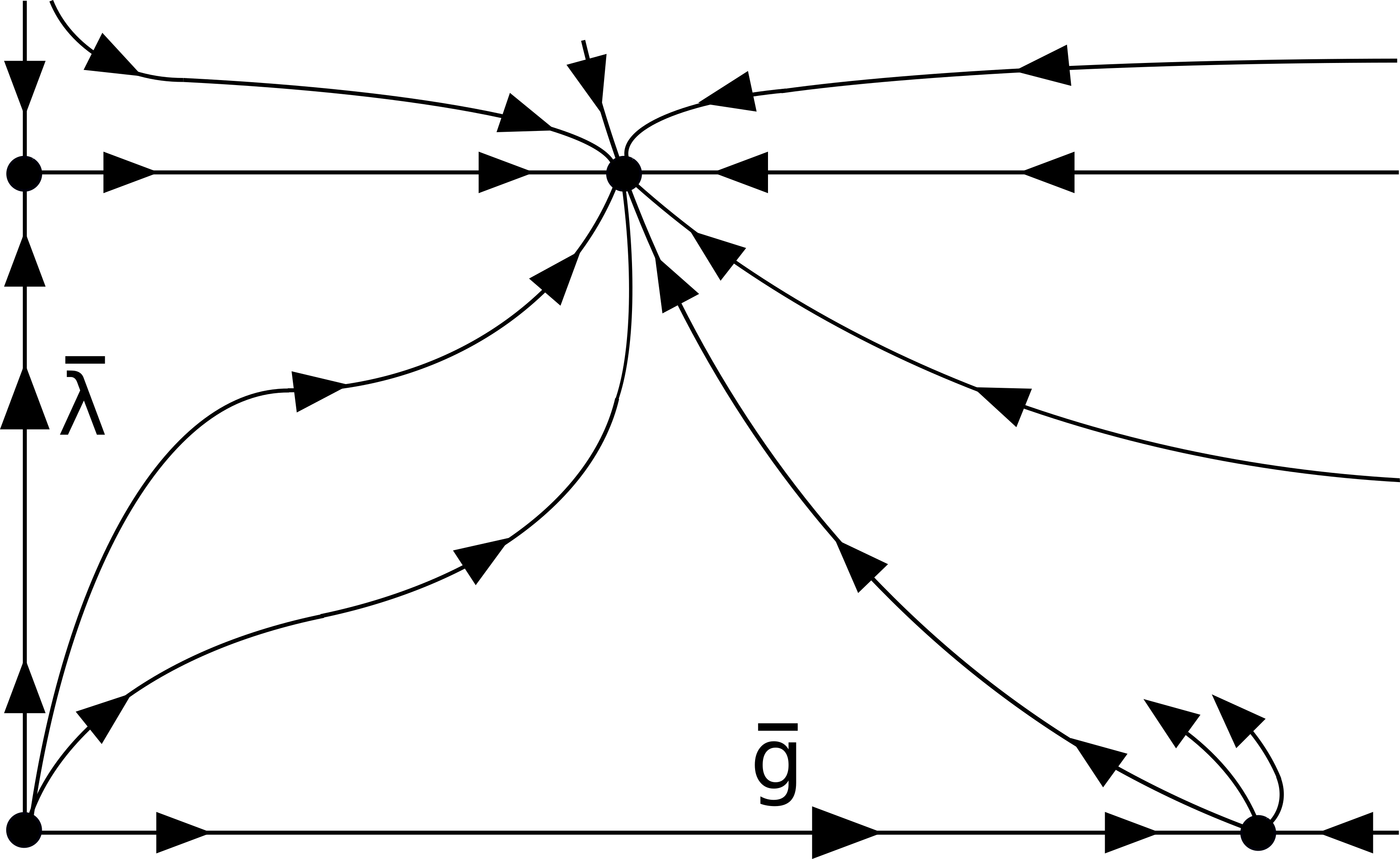}\caption{\label{fig:RG_Flow}The coupling constant flow for $\alpha>0$, $\epsilon>0$
(schematic).}
\end{figure}

\begin{lyxcode}
\begin{figure}
\centering{}\includegraphics[scale=1.4]{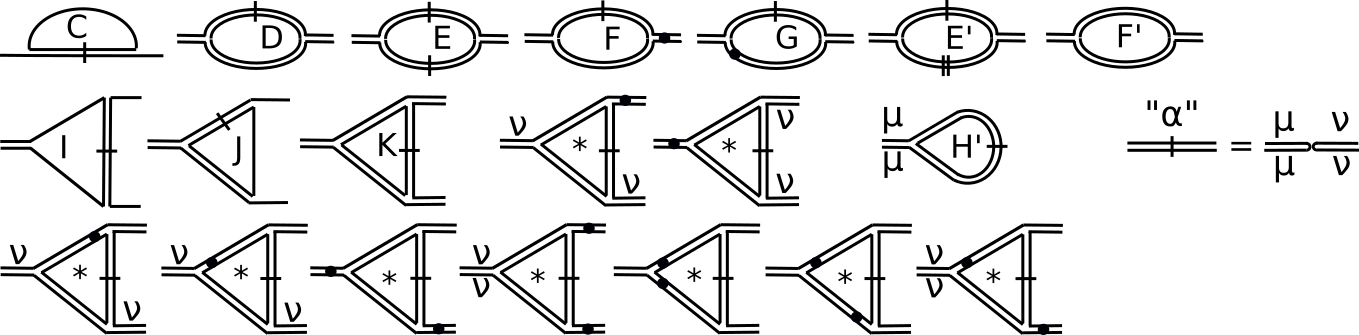}\caption{\label{fig:SpinModel_1Loop}One-loop diagrams. The dot denotes $\bar{\chi}$,
the bar denotes the $\alpha$-propagator (center right). Diagrams
marked with an asterisk (with an external dot or two internal dots
or bars) only renormalize $\lambda'$ and $\lambda''$.}
\end{figure}
\end{lyxcode}
A renormalization group calculation can be performed for $Q=\mathrm{const}$.
A one-loop calculation leads to the coupling constant flow

\begin{align}
\mathrm{d}\bar{\lambda}/\mathrm{d}l & =\bar{\lambda}\left(\tfrac{\epsilon}{2}-81\bar{\lambda}^{2}\right),\label{eq:FlowEquFinal}\\
\mathrm{d}\bar{g}/\mathrm{d}l & =\bar{g}\left(\tfrac{\epsilon}{2}-2\bar{g}^{2}-24\bar{g}\bar{\lambda}+9\bar{\lambda}^{2}\right),\nonumber 
\end{align}
where $\epsilon=8-d$ and $\bar{\lambda}=\lambda\sqrt{\alpha K_{d}}$
and $\bar{g}=g\sqrt{\alpha K_{d}}$ with $K_{d}=2^{1-d}\pi^{-d/2}/\Gamma\left(d/2\right)$.
The contributing diagrams are depicted in fig.(\ref{fig:SpinModel_1Loop}),
the coupling constant flow is shown graphically in fig.(\ref{fig:RG_Flow}).
The flow is inversion-symmetric in the $\left(\bar{g},\bar{\lambda}\right)$
plane, only the region $\bar{g}\geq0$, $\bar{\lambda}\geq0$ is of
interest. The trivial fixed point $\bar{g}=\bar{\lambda}=0$ ist unstable
below the critical dimension $8$.

At the fixed point $\bar{g}=\tfrac{1}{2}\sqrt{\epsilon}$, $\bar{\lambda}=0$
at the bottom right of fig.(\ref{fig:RG_Flow}) the finite $\bar{g}$
couples $\chi$ and $\varphi$, but there is no $\bar{\lambda}\chi^{3}$
interaction. However, $\bar{\lambda}$ is generated from $\bar{g}$,
and the fixed point is unstable in $\bar{\lambda}$-direction.

The two fixed points with a finite $\bar{\lambda}\cong\tfrac{1}{18}\sqrt{2\epsilon}$
are identical as far as the double-strand field $\chi$ is concerned.
The flow equation (\ref{eq:FlowEquFinal}) for $\bar{\lambda}$ and
the critical exponents for $\chi$ are independent of $\bar{g}$,
also in higher orders of the loop expansion. The fixed point with
finite $\bar{g}=\tfrac{1}{6}\sqrt{2\epsilon}$ in some way thus only
describes a decoration of $\chi$ with $\varphi$. 

The critical exponents $\eta=-\tfrac{\epsilon}{9}+O\left(\epsilon^{2}\right)$
and $\nu=\tfrac{1}{2}+\tfrac{5}{36}\epsilon+O\left(\epsilon\right)^{2}$
for $\chi$ agree with the exponents of one-component branched polymers
\cite{Lubensky1979}. And in fact, the critical behavior of $\chi_{\mu\nu}$
is contained in the correlation function $\left\langle \chi_{11}\chi_{11}\right\rangle ,$
in which there only occurs the $\chi_{11}$-component of $\chi_{\mu\nu}$
- the tensor nature of $\chi$ (and double polymer twisting) plays
no role.

Branched polymers of this type have been examined extensively \cite{Lubensky1979},
the most interesting aspect is the equivalence to the Lee-Yang edge-singularity
in $d-2$ dimensions \cite{Parisi1981,Brydges2003}. The Lee-Yang
exponents have been calculated in $5$-loop order \cite{Borinsky2021}.
The physically relevant exponents are the Fisher-renormalized versions
$\nu_{H}=\beta_{H}\nu$ and $\gamma_{H}=\beta_{H}\gamma$ of $\nu$
and $\gamma$, see table (\ref{tab:Exponents_chi}) \cite{Lubensky1979,LubMcKane1981}.
For example, the gyration radius of $\chi$ grows with polymer length
$s_{\chi}$ like $s_{\chi}^{\nu_{H}}$. In three dimensions the exact
value is $\nu_{H}=\tfrac{1}{2}$ \cite{Parisi1981}. The field theory
(\ref{eq:ActionIntegral}) and the RG calculation thus have shown
that the double-strand $\chi$ belongs to this universality class,
the three-point interaction originates from the $\chi\varphi\varphi$
vertex.

In contrast, the flow equation of the single-strand field $\varphi$
depends on the double-strand field $\chi$. The critical exponents
for the $\varphi$-field at the stable fixed point in one-loop order
are
\[
\eta_{\varphi}=-\tfrac{\epsilon}{18},\qquad\nu_{\varphi}=\tfrac{1}{2}+\tfrac{\epsilon}{36},\qquad\gamma_{\varphi}=\nu_{\varphi}\left(2-2\eta_{\varphi}\right)=1+\tfrac{\epsilon}{9}.
\]
The role of the $\varphi$-polymer is examined below.%

The parameter flow (\ref{eq:FlowEquFinal}) is based on the assumption
of a positive $\alpha$. Diagram $A$ in fig.(\ref{fig:Spin_Interactions})
generates a \emph{negative} contribution to $\alpha$. This diagram
involves the $\varphi$-field. Only a sufficiently strong excluded-volume
effect ($u_{2}Q^{2}$ large in in eq.(\ref{eq:Q_Def})) generates
a positive $\alpha$.

If $\alpha$ is negative (which might occur in theta solvents), then
in the flow equation (\ref{eq:FlowEquFinal}) the signs of $\bar{\lambda}^{2}$,
$\bar{g}^{2}$ and $\bar{g}\bar{\lambda}$ in the brackets change,
and there is no stable fixed point any more. The polymers strongly
interact, but this cannot be described in detail with the field theory.

\section*{Scale invariance}

\begin{table}
\begin{tabular}{|c|c|c|c|c|c|c|}
\hline 
\noalign{\vskip0.05cm}
$d$ & $\eta$ & $\nu=\left(2+\tfrac{\eta-\epsilon}{2}\right)^{-1}$ & $\gamma=\nu\left(2-\eta\right)$ & $\beta_{H}=\frac{1}{1+\gamma}$ & $\nu_{H}=\nu\beta_{H}$ & $\gamma_{H}=\gamma\beta_{H}$\tabularnewline[0.05cm]
\noalign{\vskip\doublerulesep}
\hline 
\noalign{\vskip0.05cm}
$d$ & $-\tfrac{\epsilon}{9}-\tfrac{43}{729}\epsilon^{2}$ & $\tfrac{1}{2}+\tfrac{5}{36}\epsilon+\tfrac{67}{1458}\epsilon^{2}$ & $1+\tfrac{\epsilon}{3}+\tfrac{133}{972}\epsilon^{2}$ & $\tfrac{1}{2}-\tfrac{\epsilon}{12}-\tfrac{79}{3888}\epsilon^{2}$ & $\tfrac{1}{4}+\tfrac{\epsilon}{36}+\tfrac{29}{23328}\epsilon^{2}$ & $\tfrac{1}{2}+\tfrac{\epsilon}{12}+\tfrac{79}{3888}\epsilon^{2}$\tabularnewline[0.05cm]
\noalign{\vskip\doublerulesep}
\hline 
\noalign{\vskip0.05cm}
$3$ & $-1$ & $-1$ & $-3$ & $-\tfrac{1}{2}$ & $\tfrac{1}{2}$ & $\tfrac{3}{2}$\tabularnewline[0.05cm]
\hline 
\noalign{\vskip\doublerulesep}
\end{tabular}\caption{\label{tab:Exponents_chi}Exponents for $\chi$-field at stable fixed
point in two-loop order \cite{LubMcKane1981,Borinsky2021}, $\epsilon=8-d$,
and exact values for $d=3$. Exponents with a suffix $H$ are Fisher-renormalized. }
\end{table}
The effective coupling constants of the action integral (\ref{eq:ActionIntegral})
are $\bar{\lambda}=\sqrt{\alpha K_{d}}\lambda$ and $\bar{g}=\sqrt{\alpha K_{d}}g.$
In addition to these coupling constants the action integral contains
the dimensionful parameter $\alpha$, which is known to exactly scale
like $\alpha\sim k^{2}$. Our conventions for the critical exponents
are contained in the scaling equivalences
\[
\chi\sim k^{d/2-1+\eta_{\chi}},\:\varphi\sim k^{d/2-1+\eta_{\varphi}},\:k\sim\left(\tau_{0}-\tau_{c}\right)^{\nu_{\chi}}\sim\left(r_{0}-r_{c}\right)^{\nu_{\varphi}},
\]
Critical exponents without a suffix $H$ are not Fisher-renormalized.

The scale invariance implied by the renormalization group consists
of a scaling of length, the fields and the parameter $\alpha$. Vertex
functions $\Gamma_{...}^{\mathrm{loop}}$ calculated directly in a
loop expansion (for instance with graph matching) normally scale as
predicted by the RG. This correspondance is violated if $\Gamma$
contains an $\alpha$-factor - which is constant in a direct calculation.

An example is the vertex function $\Gamma_{\chi\chi\chi}\sim k^{\epsilon/2-1-3\eta_{\chi}}$
with three amputated external $\chi$-lines, which contains a factor
$\lambda=\alpha^{-1/2}K_{d}^{-1/2}\bar{\lambda}$. A direct calculation
thus gives $\Gamma_{\chi\chi\chi}^{\mathrm{loop}}\sim\alpha^{1/2}\Gamma_{\chi\chi\chi}\sim k^{\epsilon/2-1-3\eta_{\chi}}.$
 Other examples are

\begin{alignat}{2}
\Gamma_{\chi\chi\chi} & \sim k^{\epsilon/2-1-3\eta_{\chi}}, & \quad\Gamma_{\chi\chi\chi}^{\mathrm{loop}} & \sim k^{\epsilon/2-3\eta_{\chi}},\nonumber \\
\Gamma_{\chi\varphi\varphi} & \sim k^{\epsilon/2-1-\eta_{\chi}-2\eta_{\varphi}}, & \Gamma_{\chi\varphi\varphi}^{\mathrm{loop}} & \sim k^{\epsilon/2-\eta_{\chi}-2\eta_{\varphi}},\label{eq:GammLoopScaling}\\
\Gamma_{\chi} & \sim k^{5-\epsilon/2-\eta_{\chi}}, & \Gamma_{\chi}^{\mathrm{loop}} & \sim k^{4-\epsilon/2-\eta_{\chi}}.\nonumber 
\end{alignat}
This allows to determine the exponent $\beta_{H}$ of the equation
of state (\ref{eq:EqOfState}), which also determines the Fisher-renormalization
of the critical exponents.

\section*{Equation of state}

If the $\chi$ field ist shifted like $\chi_{\mu\nu}\rightarrow\delta_{\mu\nu}Q\left(\tau_{0}\right)+\chi_{\mu\nu}$,
then the action integral takes the form (\ref{eq:Q_Def}). One requires
that for $Q\left(\tau_{0}\right)\cong Q_{c}+a\tau-A\tau^{\beta_{H}}$
the leading coefficient of $\bar{\chi}$ in the effective potential
$\Gamma$ vanishes in any order of the loop expansion. Inserting $Q\left(\tau_{0}\right)$
into eq.(\ref{eq:Q_Def}) leads to
\begin{align}
\left[...\right]+\Gamma_{\chi}^{\mathrm{loop}} & =\left(-H-\tau_{c}Q_{c}+3\lambda Q_{c}^{2}\right)+A\left(\tau_{c}-6\lambda Q_{c}\right)\tau^{\beta_{H}}\label{eq:Gamma1_eq_0}\\
 & \qquad-\left(Q_{c}+a\tau_{c}-6\lambda aQ_{c}\right)\tau+\left(A-6\lambda aA\right)\tau^{1+\beta_{H}}+3\lambda A^{2}\tau^{2\beta_{H}}+\Gamma_{\chi}^{\mathrm{loop}}=0.\nonumber 
\end{align}
The $1$-point vertex function $\Gamma_{\chi}^{\mathrm{loop}}=\mathrm{a_{0}}+a_{1}\tau^{\beta_{H}}+a_{2}\tau^{2\beta_{H}}+a_{3}\tau^{\left(d/2-\eta_{\chi}\right)\nu_{\chi}\beta_{H}}+$
... contains regular and singular contributions in the Fisher-renormalized
variable $\tau^{\beta_{H}}$, see eq.(\ref{eq:GammLoopScaling}).
The $O\left(\tau^{0}\right)$ and $O\left(\tau^{\beta}\right)$ parts
of eq.(\ref{eq:Gamma1_eq_0}) vanish with an appropriate choice of
$\tau_{c}$ and $Q_{c}$.

The $O\left(\tau\right)$-term of eq.(\ref{eq:Gamma1_eq_0}) only
can vanish for \cite{Lubensky1979}
\begin{equation}
\beta_{H}=\frac{1}{\left(d/2-\eta_{\chi}\right)\nu_{\chi}}.\label{eq:BetaH_exponent}
\end{equation}
It can be shown $\beta_{H}=1/\left(1+\gamma\right)$ \cite{LubMcKane1981}.
The critical exponents of the $\chi$-field are listed in table (\ref{tab:Exponents_chi}).
The condition $\beta_{H}<1$ for Fisher-renormalization is fullfilled.
The value $\beta_{H}^{d=3}=-\tfrac{1}{2}$ for $d=3$ is known exactly,
and thus $Q^{d=3}\sim\tau^{-1/2}$ in the vicinity of the fixed points
with $\bar{\lambda}>0$.

\section*{Polymer lengths}

\begin{figure}
\centering{}\includegraphics[scale=1.33]{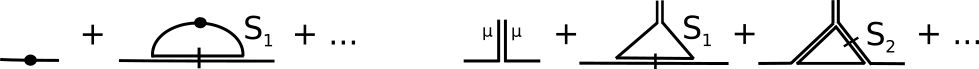}\caption{\label{fig:DgrPolymerLengths}Polymer lengths, up to one loop. Self
energies of external lines are not shown.}
\end{figure}
The critical point of the double-strand polymer $\chi$, as shown
above, is independent of the single-strand $\varphi$. The critical
exponents are the exponents of the Lee-Yang edge singularity in $d-2$
dimensions, with an additional Fisher-renormalization.

In contrast, the single-strand polymer $\varphi$ depends on the double-strand
polymer $\chi$. A realistic physical situation is one single-strand
polymer with (originally) length $s$. Within the model (\ref{eq:ActionIntegral})
this can be achieved with a weak spatially constant external field
in $\mu=1$-direction. The physics is contained in the correlation
function $\left\langle \varphi_{1}\varphi_{1}\right\rangle _{k=0}=\left\langle \varphi_{1}\left(0\right)\int\mathrm{d}^{d}x\varphi_{1}\left(x\right)\right\rangle $.
The polymer lengths $s_{\varphi}$ and $s_{\chi}$ of the polymer
types are determined by $r_{0}$ and $\tau_{0}$, which in $e^{-S}$
act like negative chemical potentials. In a real experiment, however,
the total length $s=s_{\varphi}+2s_{\chi}$ is given, and effectively
only one parameter (temperature, composition of solution, ...) can
be varied. We now write $r=r_{0}-r_{c}$ and $\tau=\tau_{0}-\tau_{c}$. 

The amounts (or lengths) of the polymer types are (fig.(\ref{fig:DgrPolymerLengths}))
\begin{align}
s_{\varphi} & =\left\langle \varphi_{1}\varphi_{1}\tfrac{1}{2}\int\mathrm{d}^{d}x\sum\varphi_{\mu}^{2}\right\rangle /\left\langle \varphi_{1}\varphi_{1}\right\rangle =\left\langle \varphi_{1}\varphi_{1}\right\rangle \Gamma_{\varphi^{2}\varphi\varphi}\sim-2\frac{\mathrm{d}}{\mathrm{d}r}\ln\left\langle e^{S}\varphi_{1}\varphi_{1}\right\rangle \sim\tfrac{1}{r},\label{eq:Expr_sPhi}\\
s_{\chi} & =\left\langle \varphi_{1}\varphi_{1}\int\mathrm{d}^{d}x\left(2Q\bar{\chi}+\sum\chi_{\mu\nu}^{2}\right)\right\rangle /\left\langle \varphi_{1}\varphi_{1}\right\rangle =2\left\langle \varphi_{1}\varphi_{1}\right\rangle \left\langle \chi_{11}\chi_{11}\right\rangle Q\Gamma_{\chi\varphi\varphi}^{\mathrm{loop}}.\label{eq:Expr_sChi}
\end{align}
The length $s_{\varphi}$ contains $\Gamma_{\varphi^{2}\varphi\varphi}$,
a two-point vertex function with a $\varphi^{2}$ insertion. The length
$s_{\chi}$ contains a contribution proportional to $Q$ because of
redefinition (\ref{eq:Q_Def}), $\chi_{\mu\nu}^{2}$ is negligible.
The vertex function $\Gamma_{\chi\varphi\varphi}^{\mathrm{loop}}$
is defined in eq.(\ref{eq:GammLoopScaling}). In Fisher-renormalized
tree-approximation
\begin{align*}
s_{\varphi} & =1/r,\qquad s_{\chi}=aQ/\left(r\tau^{1/2}\right),
\end{align*}
where $a$ is a constant. The ratio is $s_{\varphi}/s_{\chi}=\tau^{1/2}/\left(aQ\right)$.
With the constraint $s=s_{\varphi}+2s_{\chi}$ it follows
\begin{align*}
s_{\varphi} & =\frac{s}{1+2s_{\chi}/s_{\varphi}}\cong\tfrac{s}{2aQ}\tau^{1/2},\\
s_{\chi} & =\frac{s}{2+s_{\varphi}/s_{\chi}}\cong\tfrac{1}{2}s\left(1-\tfrac{\text{1}}{2aQ}\tau^{1/2}\right).
\end{align*}
In the final stage of the condensation $s_{\varphi}$ thus decreases
like $\tau^{1/2}.$

In general, if $\tau$ and $r$ are small then near a fixed point
the vertex functions scale like $\Gamma_{\varphi^{2}\varphi\varphi}\sim k^{2-1/\nu_{\varphi}-2\eta_{\varphi}}$
and $\Gamma_{\chi\varphi\varphi}^{\mathrm{loop}}\sim k^{\epsilon/2-\eta_{\chi}-2\eta_{\varphi}}$
where $k\sim\tau^{\nu_{H}}\sim r^{\nu_{\varphi}}$. The variable $\tau$
drops out from the vertex functions and from $\left\langle \varphi\varphi\right\rangle $
for $\tau\rightarrow0$. From scaling arguments $\left\langle \varphi\varphi\right\rangle \sim r^{-\gamma_{\varphi}}$.
The $\chi$-propagator $\left\langle \chi\chi\right\rangle \sim\tau^{-\gamma_{H}}$
only depends on $\tau$. This gives for $\tau\rightarrow0$
\begin{align*}
s_{\varphi} & \sim r^{-\gamma_{\varphi}}r^{\nu_{\varphi}\left(2-1/\nu_{\varphi}-2\eta_{\varphi}\right)}\sim r^{-1},\\
s_{\chi} & \sim(Q_{c}+A\tau^{\beta_{H}})\tau^{-\gamma_{H}}r^{-\beta_{H}\nu_{\varphi}/\nu_{H}},
\end{align*}
where the second line the hyperscaling equation $\beta_{H}/\nu_{H}=2-\tfrac{\epsilon}{2}+\eta_{\chi}$
has been used \cite{LubMcKane1981}.

We now assume $\beta_{H}>0$. In the final stage of the condensation
$s_{\chi}$ is constant, and thus $r^{-\beta_{H}\nu_{\varphi}/\nu_{H}}\sim\tau^{\gamma_{H}}.$
It follows
\[
s_{\varphi}\sim r^{-1}\sim\tau^{z},\qquad\mathrm{with}\;z=\frac{\gamma_{H}\nu_{H}}{\beta_{H}\nu_{\varphi}}=\tfrac{1}{2}+\tfrac{7}{36}\epsilon+...
\]
This agrees with the mean-field result for $\epsilon=0$. 

If $\beta_{H}<0$, which is the case in $d=3$ where $\nu_{H}=\tfrac{1}{2}$,
$\gamma_{H}=\tfrac{3}{2}$ and $\beta_{H}=-\tfrac{1}{2}$, then there
is no reasonable solution. The exponent $z$ diverges if $\beta_{H}$
approaches zero from above, and one might conclude that $s_{\varphi}$
suddenly drops to zero at some $\tau$-value when $\tau$ is diminished.
But the details of the condensation of a single strand to a double
strand in three dimensions are not understood.

\section*{Conclusions}

Our original starting point was a similar lattice model and a similar
field theory with length variables instead of $O\left(n\right)$ indices
\cite{Dengler2020}. This model contains directed polymers and can
describe the case where aligned and anti-aligned polymer strands form
double polymers with different binding energies. It also would allow
to examine the case where the binding energy $\tau\left(s,s'\right)$
depends on the two length variables $s$ and $s'$ at the binding
site. 

Within the simpler $O\left(n\right)$-model examined here it becomes
clear, that there is a close relationship to branched polymers consisting
of one type of polymer. The double-strand polymer acquires a three-point
interaction, decouples from the single-strand polymer near the critical
point, and thus becomes a branched polymer of the conventional type.

We have reproduced Fisher-renormalization of the critical exponents
and the equation of state in a more direct way. The critical exponents
in three dimensions are known exactly, and it would be of interest
to compare the predictions with experiment. For instance, the radius
of gyration for the $\chi$-field should grow like $s_{\chi}^{1/2}$
with polymer length $s_{\chi}$. Under most conditions, RNA strands
in a solvent are far from ideal, and one needs long polymers to reach
a scaling limit. But in principle this could be tested with RNA with
a base sequence like ATATA...

On the other hand, the single-strand polymer is somewhat ephemeral
near the critical point. It is affected by the double-strand polymer,
but disappears at the critical point, when the condensation to a double-strand
is complete. Critical exponents can be calculated in an expansion
around $d=8$, but an extrapolation to three dimensions is difficult.
And it appears that the connection between branched polymers and the
Lee-Yang edge singularity in $d-2$ dimensions \cite{Parisi1981,Brydges2003}
cannot be extended to the single-strand-component of the model. 

\section*{Appendix}

\subsection*{Appendix A: Average over spin variables}

\label{subsec:Appendix_A}The analytic expression for the partition
sum makes use of an average over components $\sigma_{i}$ of an $N$-dimensional
vector $\left(\sigma_{1},\sigma_{2},...\sigma_{N}\right)$ with weight
$P_{N}\left(\boldsymbol{\sigma}\right)\mathrm{d}^{N}\sigma=\pi^{-N/2}\tfrac{1}{2}\Gamma\left(\tfrac{N}{2}\right)\delta\left(\left|\boldsymbol{\sigma}\right|/\sqrt{N}-1\right)\mathrm{d}^{N}\sigma$.
One finds

\[
\left\langle \sigma_{1}^{m_{1}}...\sigma_{k}^{m_{k}}\right\rangle _{N}=\int P_{N}\left(\boldsymbol{\sigma}\right)\sigma_{1}^{m_{1}}...\sigma_{k}^{m_{k}}\mathrm{d}^{N}\sigma=\frac{\Gamma\left(\tfrac{N}{2}\right)N^{\left(m_{1}+...m_{k}\right)/2}}{\Gamma\left(\tfrac{N+m_{1}+...m_{k}}{2}\right)2^{\left(m_{1}+...m_{k}\right)/2}}\left\langle \sigma_{1}^{m_{1}}..\sigma_{k}^{m_{k}}\right\rangle _{\mathrm{\mathrm{G}}},
\]
where $\left\langle ...\right\rangle _{\mathrm{\mathrm{G}}}$ is the
normalized average over the gaussian function $\exp\left(-\tfrac{1}{2}\boldsymbol{\sigma}^{2}\right).$
The gaussian averages can be calculated by pairwise contraction of
the $\sigma_{i}$-factors, and produce a value of order $1$. This
gives $\left\langle 1\right\rangle _{N}=1$ and $\left\langle x_{i}x_{j}\right\rangle =\delta_{ij}$.
Averages are analytic in $N$, and expectation values of $m>2$ $\sigma$-factors
vanish like $N^{\left(m-2\right)/2}$ in the limit $N\rightarrow0$.

\subsection*{Appendix B: Interaction at the level of the lattice model }

\label{subsec:Appendix_B}The ansatz for the interaction at a given
lattice point is $f=f_{1}+bf_{2}+cf_{3}$, where
\[
f_{1}=\tfrac{\left(N+2\right)\left(N+4\right)}{2N^{2}}\sum_{\mu\nu}S^{\mu\nu}s^{\mu}s^{\nu},\qquad f_{2}=\sum_{\mu\nu}S^{\mu\mu}s^{\nu}s^{\nu}\qquad f_{3}=\sum_{\mu}S^{\mu\mu}.
\]
One requires $\left\langle S^{\alpha\beta}s^{\rho}s^{\tau}f\right\rangle =\delta_{\alpha\rho}\delta_{\beta\tau}+\delta_{\alpha\tau}\delta_{\beta\rho}$
and $\left\langle S^{\alpha\beta}f\right\rangle =0.$ This is possible
only with factors 1/$N$. It follows
\[
\begin{array}{lcl}
\left\langle S^{\alpha\beta}s^{\rho}s^{\tau}f_{1}\right\rangle =\delta_{\alpha\rho}\delta_{\beta\tau}+\delta_{\alpha\tau}\delta_{\beta\rho}+\delta_{\alpha\beta}\delta_{\rho\tau}, &  & \left\langle S^{\alpha\beta}f_{1}\right\rangle =\tfrac{N+4}{N}\delta_{\alpha\beta},\\
\left\langle S^{\alpha\beta}s^{\rho}s^{\tau}f_{2}\right\rangle =\tfrac{2N^{2}\left(n+2\right)}{\left(N+2\right)\left(N+4\right)}\delta_{\alpha\beta}\delta_{\rho\tau}, &  & \left\langle S^{\alpha\beta}f_{2}\right\rangle =\tfrac{2Nn}{N+2}\delta_{\alpha\beta},\\
\left\langle S^{\alpha\beta}s^{\rho}s^{\tau}f_{3}\right\rangle =\tfrac{2N}{N+2}\delta_{\alpha\beta}\delta_{\rho\tau}, &  & \left\langle S^{\alpha\beta}f_{3}\right\rangle =2\delta_{\alpha\beta}.
\end{array}
\]
The solution is
\[
b=\tfrac{\left(N+2\right)\left(N+4\right)}{2N^{2}\left(2+N-n\right)},\qquad c=-Nb.
\]

\bibliographystyle{aip}
\bibliography{DoublePolymerSpin.bib}

\end{document}